\documentclass[fleqn,twoside]{article}
\usepackage{amsmath,amssymb,bm,curves}
\usepackage{espcrc2}

\newcommand{\run}[1]{\widetilde{\alpha}_{#1}}
\newcommand{\hsp}[1]{\hspace*{#1 mm}}

\title{Decoupling heavy particles simultaneously}

\author{R.J.\ Crewther\address[Adelaide]{Department of Physics and Centre 
for the Subatomic Structure of Matter (CSSM), 
University of Adelaide, SA 5005, Australia},
S.D.\ Bass\address{High Energy Physics Group,
Institute for Experimental Physics and
Institute for Theoretical Physics, Universit\"at Innsbruck,
Technikerstrasse 25, A 6020 Innsbruck, Austria}%
\thanks{Research supported by the Austrian Science Fund (FWF) (grant M770). 
The CSSM is thanked for its hospitality.}, 
F.M.\ Steffens\address{Instituto de Fisica Teorica,
Rua Pamplona 145, 01405-900 S\~ao Paulo, SP, Brazil}%
\address{NFC\:-\:FCBEE\:-\:Universidade Presbiteriana Mackenzie,
Rua da Consola\c{c}\~ao 930, 01302-907, S\~ao Paulo, SP, Brazil}%
\thanks{Supported by FAPESP (03/10754-0), CNPq (308932/   
2003-0) and MackPesquisa.}
and
A.W.\ Thomas\addressmark[Adelaide]\address%
{Jefferson National Laboratory, 12000 Jefferson Ave.,
Newport News, VA 23606, USA}%
\thanks{Supported by the Australian Research Council and by DOE
contract DE-AC05-84ER40150, under which SURA operates Jefferson 
National Laboratory.}}   

\begin{document}

\begin{abstract}
The renormalization group is extended to cases where several heavy particles
are decoupled at the same time. This involves large logarithms which are
\emph{scale-invariant} and so \emph{cannot} be eliminated by a change of 
renormalization scheme. A set of scale-invariant running couplings, one 
for each heavy particle, is constructed without reference to intermediate 
thresholds. The entire heavy-quark correction to the axial
charge of the weak neutral current is derived to next-to-leading order, 
and checked in leading order by evaluating diagrams explicitly. The 
mechanism for cancelling contributions from the top and bottom quarks 
in the equal-mass limit is surprisingly non-trivial.

\vspace{1pc}
\end{abstract}

\maketitle

\section{SEVERAL LARGE SCALES}

When the renormalization group (RG) is applied to a problem involving 
two or more large masses or momenta, the standard procedure is to deal
with each asymptotic scale separately, starting with the largest.
However, there are many cases of practical interest where this sequential
approach will not do.  For example, the top quark, the $W^\pm$ and $Z^0$
bosons, and presumably the Higgs boson have masses in the same region,
so in situations where more than one of these particles sets the scale, 
or there are momenta of similar magnitude, it is not reasonable to suppose 
that one of the corresponding logarithms dominates the others.  Various
techniques for dealing with this situation are being investigated
\cite{smir,blum,been,denn,brod}.

Our observation \cite{BCST1,BCST2,BCST3} is that the RG specifies
simultaneous dependence on asymptotic scales 
\begin{equation}
M_1 \geqslant M_2 \geqslant \ldots \geqslant M_n \gg 
\mbox{ fixed scale }\mu\,,
\end{equation}
for limits in which their logarithms grow large together:
\begin{equation}
\ln(M_i/\mu)\!\bigm/\!\ln(M_j/\mu) = \mbox{fixed, } M_i,M_j \to \infty\,.
\label{1a}\end{equation}
The key new feature is the appearance of large logarithms
\begin{equation}
\ln(M_i/M_j) = \ln(M_i/\mu) - \ln(M_j/\mu)
\label{1d}\end{equation}
which are \emph{scale-invariant}, i.e.\ independent of the 
renormalization scale $\mu$.  This precludes any use of the standard 
method \cite{CWZ,marc} where a running coupling evolves from 
threshold to threshold and where, at each new threshold $M_j$, large 
scale-\emph{dependent} logarithms are suppressed by choosing a new scale 
$\mu \sim M_j$.

Instead, for each heavy particle $h_i$ (or large momentum $Q$), we 
construct a scale-invariant running coupling 
\begin{equation}
\alpha_{h_i} 
= \alpha_{h_i}\bigl(\ln(M_1/\mu), \ldots , \ln(M_n/\mu)\bigr)  
\end{equation}
\emph{without} reference to intermediate thresholds.  All asymptotic
dependence on $M_1, \ldots M_n$ is carried by $\{\alpha_{h_i};
i = 1, \ldots ,n\}$. Coefficients in the Taylor series in 
$\{\alpha_{h_i}\}$ are determined by matching with the result 
for the sequential limit
\begin{equation}
M_j \to \infty \mbox{ for fixed } M_{j+1}, \ldots M_n,\ j = 1,2,\ldots
\label{1c}\end{equation}

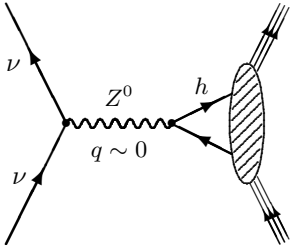
\begin{figure}[t]
\begin{center}
\setlength{\unitlength}{1mm}
\begin{picture}(44,32)(0,0)
\thicklines
\put(3,0){\vector(1,2){5}}
\put(3,0){\line(1,2){8}}
\put(11,16){\vector(-1,2){5}}
\put(11,16){\line(-1,2){8}}
\put(11,16){\circle*{1}}
\put(3.8,8){$\nu$}
\put(3,23){$\nu$}
\put(18,18.4){\makebox[0mm]{\small $Z^0$}}
\put(18,11.6){\makebox[0mm]{\small $q \sim 0$}}
\put(28,20){\small $h$}
\newcommand{\glue}%
{\curve(0,0, 0.5,0.6, 1,0)\curve(1,0, 1.5,-0.6, 2,0)
\curve(2,0, 2.5,0.6, 3,0)\curve(3,0, 3.5,-0.6, 4,0)
\curve(4,0, 4.5,0.6, 5,0)\curve(5,0, 5.5,-0.6, 6,0)
\curve(6,0, 6.5,0.6, 7,0)\curve(7,0, 7.5,-0.6, 8,0)}
\put(11,16){\glue}\put(17,16){\glue}
\put(25,16){\vector(2,1){6}}
\put(25,16){\line(2,1){8}}
\put(33,12){\vector(-2,1){5}}
\put(33,12){\line(-2,1){8}}
\put(25,16){\circle*{1}}
\thinlines
\put(35.4,24){\line(1,2){4}}\put(35.4,24){\vector(1,2){2.6}}
\put(36,23.7){\line(1,2){4}}\put(36,23.7){\vector(1,2){2.6}}
\put(36.4,23){\line(1,2){4.2}}\put(36.4,23){\vector(1,2){2.8}}
\put(35.4,8){\line(1,-2){4}}\put(39.4,0){\vector(-1,2){2.2}}
\put(36,8.3){\line(1,-2){4}}\put(40,0.3){\vector(-1,2){2.2}}
\put(36.4,9){\line(1,-2){4.2}}\put(40.4,1){\vector(-1,2){2}}
\curve(33.6,21.5, 35.5,23.4)
\curve(33.3,19.7, 36.05,22.45)
\curve(33.1,18.0, 36.4,21.3)
\curve(33.0,16.4, 36.65,20.05)
\curve(32.9,14.8, 36.8,18.7)
\curve(33.1,13.5, 36.95,17.35)
\curve(33.3,12.2, 37,15.9)
\curve(33.5,10.9, 36.95,14.35)
\curve(33.9,9.8, 36.7,12.6)
\curve(34.35,8.75, 36.5,10.9)
\linethickness{0.2mm}
\closecurve(33,20, 35,24, 37,20, 37,12, 35,8, 33,12)
\end{picture}
\end{center}
\caption{Neutrino scattering from nucleons at low momentum transfers 
         $q$. \label{fig:1}}
\end{figure}

These ideas are tested for the classic example \cite{CWZ,KM,CK} of heavy 
quarks $h$ being decoupled from the weak neutral axial-vector current
\begin{equation}
J_{\mu 5\rule{0mm}{2mm}}^{\mathrm{Weak}\rule[-1mm]{0mm}{2mm}}\, 
 =\, \frac{1}{2}\Biggl\{\sum_{q=u,c,t}\, -\, \sum_{q=d,s,b}\,\Biggr\}\,
     \bar{q}\gamma_\mu\gamma_5 q
\label{2a} \end{equation}
at low momentum transfer $q$ (Fig.~\ref{fig:1}). In particular, we
check that our results for simultaneous $t,b$ decoupling are consistent
with the sequential and $m_b \sim m_t$ limits. 

Results are presented for three-colour quantum chromodynamics (QCD) with
modified minimally subtracted ($\overline{\mbox{\small MS}}$) 
renormalization at fixed scale $\bar{\mu}$.

\begin{figure}[t]
\begin{center}
\setlength{\unitlength}{1mm}
\begin{picture}(36,24)
\put(8,12){\circle*{1}}
\put(16,4){\circle*{1}}
\put(16,20){\circle*{1}}
\put(28,4){\circle*{1}}
\put(28,20){\circle*{1}}
\put(0,11.5){$\gamma_\mu\gamma_5$} 
\put(13.5,11.5){$h$}
\put(29,11.5){$\ell$}
\thicklines
\put(16,4){\line(0,1){16}}
\put(28,4){\line(0,1){16}}
\put(8,12){\line(1,1){8}}
\put(8,12){\line(1,-1){8}}
\put(28,4){\line(2,-1){8}}
\put(28,20){\line(2,1){8}}
\newcommand{\glue}%
{\curve(0,0, 0.5,0.6, 1,0)\curve(1,0, 1.5,-0.6, 2,0)
\curve(2,0, 2.5,0.6, 3,0)\curve(3,0, 3.5,-0.6, 4,0)
\curve(4,0, 4.5,0.6, 5,0)\curve(5,0, 5.5,-0.6, 6,0)
\curve(6,0, 6.5,0.6, 7,0)\curve(7,0, 7.5,-0.6, 8,0)}
\put(16,20){\glue}
\put(20,20){\glue}
\put(16,4){\glue}
\put(20,4){\glue}
\end{picture}
\caption{The lowest-order diagrams (two of them when fermion arrows are 
         restored) responsible for mass logarithms induced by the neutral 
         current (\ref{2a}).  Gluon propagators are shown as
         wavy lines.  Straight lines denote propagators for heavy ($h$) 
         and light ($\ell$) quarks.  \label{fig:2}}
\end{center}
\end{figure}
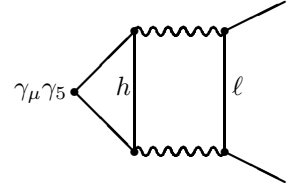

\section{DIAGRAMMATIC ANALYSIS}

In perturbation theory, mass logarithms are first produced by the 
two-loop graphs of Fig.~\ref{fig:2}. When both $t$ and $b$ are heavy, 
the result is \cite{CWZ}
\begin{align} 
\big\langle\bar{t}\gamma_\mu\gamma_5t 
  &- \bar{b}\gamma_\mu\gamma_5b\big\rangle_6 \nonumber \\
  &= \Bigl(\frac{\alpha_6}{\pi}\Bigr)^2\ln\frac{m_t}{m_b}\gamma_\mu\gamma_5  
     + O\bigl(m_{t,b}^{-2},\alpha_6^3\bigr)\,.
\label{2b} \end{align}
A subscript $F$ refers to the number of flavours ($F=6$ in this case); 
thus $\alpha_6 = g^2_6/4\pi^2$ is the $\overline{\mbox{\small MS}}$
renormalized strong coupling for 6-flavour QCD. Eq.~(\ref{2b})
follows directly from Adler's two-loop renormalization $Z$-factor 
\cite{Adler} for the gauge-invariant axial-vector current in quantum 
electrodynamics (QED). 

At this two-loop level, the manner in which $m_t$ and $m_b$ are taken 
to infinity is immaterial:  the limits can be taken separately or
together, and $\alpha_6$, $\alpha_5$, and  $\alpha_4$ need not 
be distinguished.

\begin{figure*}[t]
\begin{center}
\setlength{\unitlength}{1.2mm}
\begin{picture}(104.5,24)
\thicklines
\linethickness{0.28mm}
\newcommand{\glue}%
{\curve(0,0, 0.5,0.6, 1,0)\curve(1,0, 1.5,-0.6, 2,0)
\curve(2,0, 2.5,0.6, 3,0)\curve(3,0, 3.5,-0.6, 4,0)
\curve(4,0, 4.5,0.6, 5,0)\curve(5,0, 5.5,-0.6, 6,0)}
\newcommand{\gleg}%
{\curve(0,0, 0.5,0.6, 1,0)\curve(1,0, 1.5,-0.6, 2,0)
\curve(2,0, 2.5,0.6, 3,0)\curve(3,0, 3.5,-0.6, 4,0)}
\newcommand{\gcircle}%
{\curve(0.8,0.6, 1.3,0.8, 1.4,1.4)\curve(1.4,1.4, 1.65,2.25, 2.4,1.8)
\curve(0.8,-0.6, 1.3,-0.8, 1.4,-1.4)\curve(1.4,-1.4, 1.65,-2.25, 2.4,-1.8)
\curve(2.4,1.8, 3,1.5, 3.6,1.8)\curve(3.6,1.8, 4.35,2.25, 4.6,1.4)
\curve(2.4,-1.8, 3,-1.5, 3.6,-1.8)\curve(3.6,-1.8, 4.35,-2.25, 4.6,-1.4)
\curve(4.6,1.4, 4.7,0.8, 5.2,0.6)\curve(5.2,0.6, 6,0, 5.2,-0.6)
\curve(4.6,-1.4, 4.7,-0.8, 5.2,-0.6)\curve(0.8,0.6, 0,0, 0.8,-0.6)}
\newcommand{\gloop}{\gleg \put(4,0){\gcircle} \put(10,0){\gleg}}
\newcommand{\cterm}{\gleg \curve(4,0, 5.6,0) 
                     \put(3.95,-0.93){\textsf{X}}\put(5.6,0){\gleg}}
\newcommand{\qloop}%
{\curve(0,0, 0.5,0.6, 1,0)\curve(1,0, 1.5,-0.6, 2,0)
\curve(2,0, 2.5,0.6, 3,0)\curve(3,0, 3.5,-0.6, 4,0)
\put(5.45,-4.25){\small $q$}\closecurve(4,0, 6,2, 8,0, 6,-2)
\curve(8,0, 8.5,0.6, 9,0)\curve(9,0, 9.5,-0.6, 10,0)
\curve(10,0, 10.5,0.6, 11,0)\curve(11,0, 11.5,-0.6, 12,0)
{\thinlines \put(5.7,1.95){\vector(1,0){1}}\put(6.4,-1.95){\vector(-1,0){1}}}}
\put(16,4){\glue}
\put(16,20){\glue}
\put(8,12){\circle*{1}}
\put(16,4){\circle*{1}}
\put(16,20){\circle*{1}}
\put(16,4){\line(0,1){16}}
\put(8,12){\line(1,1){8}}
\put(8,12){\line(1,-1){8}}
\put(1.2,11){$2J^Z_{\mu 5}$}
\put(13.5,11){\small $h$}
\newcommand{\series}{\small $\displaystyle 
\biggl\{\raisebox{-0.4mm}{\large 1}\, 
-\ \raisebox{1mm}{\gloop}\hspace{16.8mm}\
-\ \sum_{\raisebox{0.4mm}{\scriptsize $q$}}\raisebox{1mm}{\qloop} 
\hspace{14.2mm}\ -\ \raisebox{1mm}{\cterm}\hspace{12.3mm}\biggr\}%
\raisebox{4mm}{\scriptsize \hspace{-0.8mm}$-1$}$}
\put(23,19){\series}
\put(23,3){\series}
\put(87,4){\glue\curve(6,0, 6.5,0.6, 7,0)\curve(7,0, 7.5,-0.6, 8,0)}
\put(87,20){\glue\curve(6,0, 6.5,0.6, 7,0)\curve(7,0, 7.5,-0.6, 8,0)}
\put(95,4){\line(0,1){16}}
\put(95,4){\line(2,-1){8}}
\put(95,20){\line(2,1){8}}
\put(95,4){\circle*{1}}
\put(95,20){\circle*{1}}
\put(96.3,11){\small $\ell$}
\end{picture}
\caption{Graphs responsible for leading-order mass logarithms in 
ghost-free gauges. Each gluon propagator is dressed with one-loop gluon 
and quark bubbles (all six flavours $q$) plus counterterms; these 
insertions sum to a geometric series, as shown.  Tadpoles and other 
one-point gluon insertions are renormalized to zero.  Not shown are 
heavy-quark corrections to the light-quark propagator which produce
finite corrections to make the residual singlet axial-vector current 
scale-invariant, as in Eq.~(\ref{inv}). 
\label{fig:3}}
\end{center}
\end{figure*}
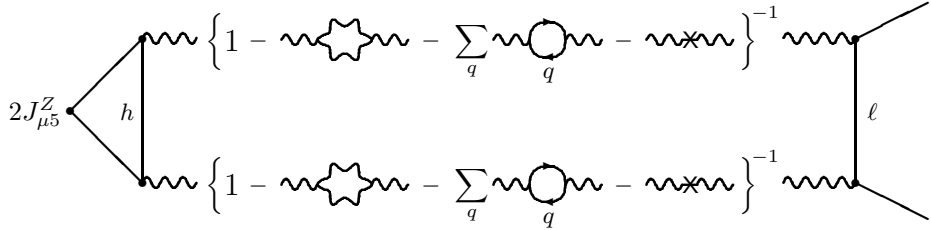

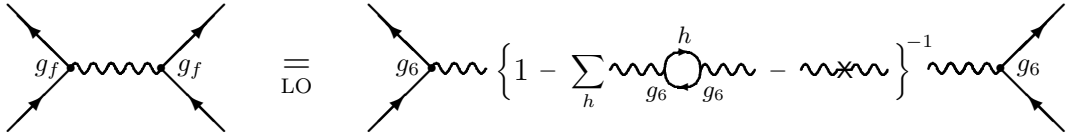
\begin{figure*}[t]
\begin{center}
\setlength{\unitlength}{1.2mm}
\begin{picture}(117,14)
\thicklines
\linethickness{0.28mm}
\newcommand{\glue}%
{\curve(0,0, 0.5,0.6, 1,0)\curve(1,0, 1.5,-0.6, 2,0)
\curve(2,0, 2.5,0.6, 3,0)\curve(3,0, 3.5,-0.6, 4,0)
\curve(4,0, 4.5,0.6, 5,0)\curve(5,0, 5.5,-0.6, 6,0)}
\newcommand{\gleg}%
{\curve(0,0, 0.5,0.6, 1,0)\curve(1,0, 1.5,-0.6, 2,0)
\curve(2,0, 2.5,0.6, 3,0)\curve(3,0, 3.5,-0.6, 4,0)}
\newcommand{\cterm}%
{\gleg \curve(4,0, 5.6,0)\put(3.95,-0.93){\textsf{X}}\put(5.6,0){\gleg}}
\newcommand{\hloop}%
{\put(0,0){\glue}
\put(7.45,3){\small $h$}
\put(3.9,-3){$g_6$}\put(10.2,-3){$g_6$}
\closecurve(6,0, 8,2, 10,0, 8,-2)
{\thinlines \put(7.7,1.95){\vector(1,0){1}}\put(8.4,-1.95){\vector(-1,0){1}}}
\put(10,0){\glue}}
\newcommand{\series}{\small $\displaystyle 
\biggl\{\raisebox{-0.4mm}{\large 1}\, 
-\ \sum_{\raisebox{0.4mm}{\scriptsize $h$}}\raisebox{1mm}{\hloop} 
\hspace{19.2mm}\ -\ \raisebox{1mm}{\cterm}\hspace{12.3mm}\biggr\}%
\raisebox{4mm}{\scriptsize \hspace{-0.8mm}$-1$}$}
\newcommand{\vertexA}%
{\put(0,0){\circle*{1}}
\put(0,0){\line(-1,1){7}}\put(0,0){\line(-1,-1){7}}
\put(0,0){\vector(-1,1){5}}\put(-5,-5){\vector(1,1){2}}}  
\newcommand{\vertexB}%
{\put(0,0){\circle*{1}}
\put(0,0){\line(1,1){7}}\put(0,0){\line(1,-1){7}}
\put(0,0){\vector(1,1){5}}\put(5,-5){\vector(-1,1){2}}}  
\put(3,6.7){$g_f$}
\put(7,7){\vertexA}
\put(7,7){\glue}\put(11,7){\glue}
\put(17,7){\vertexB}
\put(18.8,6.7){$g_f$}
\put(32,6.5){\makebox[0mm]{\Large $=$}}
\put(32,4.3){\makebox[0mm]{\footnotesize LO}}
\put(43,6.7){$g_6$}
\put(47,7){\vertexA}
\put(47,7){\glue}
\put(54,6){\series}
\put(102,7){\glue}\put(104,7){\glue}
\put(110,7){\vertexB}
\put(111.8,6.7){$g_6$}
\end{picture}
\caption{Leading-order matching condition for the $f$-flavour 
coupling $\alpha_f = g_f^2/4\pi$ produced by decoupling $6-f$ heavy 
quarks $h$ from the six-flavour theory with coupling 
$\alpha_6 = g_6^2/4\pi$.
\label{fig:4}}
\end{center}
\end{figure*}

However, we must be more careful when summing all leading-order (LO) 
logarithms 
\[ \sim \alpha_F^{r+s+1}\ln^r(m_t/\bar{\mu})\ln^s(m_b/\bar{\mu}) 
   \mbox{, integer } r,s \geqslant 0\,.  \]
For example, if $m_t$ and $m_b$ tend to $\infty$ together, 
$\alpha_5$ \emph{cannot} enter the analysis, and the coupling $\alpha_4$ of 
the residual $F=4$ theory must be held fixed (\emph{not} $\alpha_6$).

Two types of diagrams have to be taken into account. First, there are the 
diagrams in Fig.~\ref{fig:3} which produce LO logarithms controlled by 
the RG; these depend on the strong coupling $\alpha_6$ of the 
\emph{original} 6-flavour theory. Then the diagrams of 
Fig.~\ref{fig:4} are needed to obtain the LO matching condition 
expressing $\alpha_6$ in terms of the coupling $\alpha_f$ of the
\emph{residual} theory with $f$ light flavours.

\subsection{Preliminary Calculation}

First we check the LO amplitude $\Gamma_{\mu 5}^{(t)}$ 
for the top quark $t$ to decouple from $\bar{t}\gamma_\mu\gamma_5t - 
\bar{b}\gamma_\mu\gamma_5b$. In this case, the top mass $m_t$ acts as 
a regulator for the singlet component
\[ \frac{1}{5}\sum_{q=\ell} \bar{q}\gamma_\mu\gamma_5 q ,\ 
\ell = u,d,s,c,b \]
of the current $\bar{b}\gamma_\mu\gamma_5 b$ in the residual 5-flavour 
theory, with a renormalizing $Z$-factor produced by the diagrams of 
Fig.~\ref{fig:3} in LO approximation. All dependence on $\alpha_6$ 
has to be eliminated in favour of $\alpha_5$, the quantity held fixed 
as $m_t\to\infty$. So we apply Fig.~\ref{fig:4} with $f=5$,
\begin{equation}
\alpha_5 \underset{\mathrm{LO}}{=} 
   \alpha_6\bigm/\bigl\{1 - \bigl(\alpha_6/3\pi\bigr)
             \ln\bigl(m_t/\bar{\mu}\bigr)\bigr\}\,, 
\end{equation}
and invert to obtain the matching condition
\begin{equation}
\alpha_6 \underset{\mathrm{LO}}{=}
   \alpha_5\bigm/\bigl\{1 + \bigl(\alpha_5/3\pi\bigr)
             \ln\bigl(m_t/\bar{\mu}\bigr)\bigr\}\,.
\end{equation}

After some work on the diagrams of Fig.~\ref{fig:3}, we obtain an 
asymptotic formula
\begin{equation}
\Gamma_{\mu 5}
\sim - \langle\bar{b}\gamma_\mu\gamma_5b\rangle_5 -
        (6/23)^2 \gamma_\mu\gamma_5 \!\bigm/\!
         \ln(m_t/\bar{\mu})
\label{f=5}\end{equation}
in complete agreement with the result of standard RG analysis \cite{CWZ,CK}. 
The number 23 comes from the gluon self-energy insertions, i.e. 
from the one-loop factor $33-2f$ in the beta function 
\begin{align}
\beta_f(x)    
= - \frac{x^2}{6\pi}(33-2f) - \frac{x^3}{12\pi^2}&(153 - 19f)  \nonumber \\
    &+ O(x^4) 
\end{align}
of the \emph{residual} $f=5$ theory.  The bare vertex $\gamma_\mu\gamma_5$ 
corresponds to the gauge-invariant singlet operator normalised 
scale-invariantly \cite{CK,BCST1}, 
\begin{align}
\Bigl(J_{\mu 5}^\mathrm{inv}\Bigr)_f 
&= \sum_{q=\ell}\bigl(\bar{q}\gamma_\mu\gamma_5q\bigr)_f^{\mathrm{g.\;inv}}
 \exp\int_0^{\alpha_f}\!\!dx\,\frac{\gamma_f(x)}{\beta_f(x)}\ ,
\nonumber \\
  \ell &= \mbox{ all $f$ light flavours}\,,
\label{inv}\end{align}
where $f=5$ in Eq.~(\ref{f=5}), and where \cite{Larin,CK}
\begin{equation}
\gamma_f(x)
= \frac{x^2}{\pi^2}f + \frac{x^3}{36\pi^3}
            \bigl(177\!-\!2f\bigr)f + O(x^4)
\end{equation}
is the Callan-Symanzik function for the anomalous dimension of the 
singlet current.

\subsection{A Paradox?}

Suppose that both $t$ and $b$ are decoupled from 
$\bar{t}\gamma_\mu\gamma_5t - \bar{b}\gamma_\mu\gamma_5b$.
The residual theory has $f=4$ flavours, and all diagrams in 
Fig.~\ref{fig:3} shrink to the bare coupling $\gamma_\mu\gamma_5$, 
which corresponds to the $f=4$ version of the invariant current (\ref{inv}).

This case is interesting because of the role played by the equal-mass 
limit $m_b = m_t$, for which the answer to any order must vanish.  To cancel
the $t$ term in (\ref{f=5}) at $m_b = m_t$, a \emph{five}-flavour 
$b$ contribution has been suggested \cite{CWZ,CK}:
\begin{align}
\Gamma_{\mu 5}^\mathrm{cancel} &\sim (6/23)^2 \gamma_\mu\gamma_5    
\nonumber \\
&\times\Bigl\{\bigl(\ln(m_b/\bar{\mu})\bigr)^{-1}   
            - \bigl(\ln(m_t/\bar{\mu})\bigr)^{-1} \Bigr\}\,.
\label{cancel}\end{align}
However \emph{sequential} decoupling of $t$ and then $b$ does not reproduce 
this result. When $b$ is decoupled directly from 
$\langle\bar{b}\gamma_\mu\gamma_5b\rangle_5$ in Eq.~(\ref{f=5}), 
standard RG theory gives \cite{BCST1}
\begin{align}
\Gamma_{\mu 5}^\mathrm{seq} &\sim (36/23) \gamma_\mu\gamma_5     \nonumber \\
\times &\Bigl\{\bigl(25\ln(m_b/\bar{\mu})\bigr)^{-1} 
            - \bigl(23\ln(m_t/\bar{\mu})\bigr)^{-1} \Bigr\}\,.
\label{seq}\end{align}
A pure five-flavour result is \emph{not} obtained because the residual theory 
has \emph{four} flavours -- hence the factor 25 ($=33-2f$ for $f=4$) in 
the $b$ term.

In fact, there is no paradox here:  the limits $m_t \sim m_b$ and
$m_t \gg m_b$ are extreme cases of the general limit (\ref{1a}).  
To understand what is happening, it is necessary to keep track of 
dependence on the scale-invariant logarithm $\ln(m_t/m_b)$.

\subsection{Simultaneous $t,b$ Decoupling}

The leading logarithms of diagrams in Fig.~\ref{fig:3} do not depend
on external momenta or light-quark masses $m_\ell$, so we set them equal
to zero. The amplitude can be written as a Euclidean integral over 
the momentum $Q$ carried by each dressed gluon propagator, 
with self-energy insertions $\Pi(-Q^2)$:  
\begin{align}
\Gamma_{\mu 5}^{t,b} \underset{\mathrm{LO}}{=}&\
  \frac{\alpha_6^2}{2\pi^2}\gamma_\mu\gamma_5 \int^1_0\!ds\,
  \int^\infty_{m_b^2}\frac{dQ^2}{Q^2\bigl\{1 - \Pi(-Q^2)\bigr\}^2} 
\nonumber \\
  &\times\Bigl[\bigl\{1 + s(1-s)\bigl(Q/m_h\bigr)^2\bigr\}^{-1}\Bigr]_{h=b}^t
 \,.
\end{align}
The gluon integral has been restricted to the relevant asymptotic region 
$|Q| \gtrsim m_b$.  It can be further restricted by noting that, as $t$
and $b$ decouple, the factor in square brackets tends to 1 for 
$m_b \lesssim |Q| \lesssim m_t$ and is negligible otherwise:
\begin{equation}
\Gamma_{\mu 5}^{t,b}
\underset{\mathrm{LO}}{=}
  \frac{\alpha_6^2}{2\pi^2}\gamma_\mu\gamma_5
  \int^{m_t^2}_{m_b^2}\frac{dQ^2}{Q^2}
     \frac{1}{\bigl(1 - \Pi(-Q^2)\bigr)^2}\:.
\end{equation}
In this $Q$-range, the gluon self-energy amplitude is
\begin{equation}
\Pi(-Q^2) \underset{\mathrm{LO}}{=}
 - \frac{23\alpha_6}{12\pi}\ln\frac{Q^2}{\bar{\mu}^2} +
    \frac{\alpha_6}{6\pi}\ln\frac{m_t^2}{\bar{\mu}^2}\,,
\end{equation}
so we are left with a simple logarithmic integral:
\begin{align}
\Gamma_{\mu 5}^{t,b}
\underset{\mathrm{LO}}{=}\ &\frac{6\alpha_6}{23\pi}\gamma_\mu\gamma_5
\biggl[-\ \Bigl\{1 + \frac{\alpha_6}{6\pi}
         \Bigl(21\ln\frac{m_t}{\bar{\mu}}\Bigr)\Bigr\}^{-1}  \nonumber \\
+\ &\Bigl\{1 + \frac{\alpha_6}{6\pi}\Bigl(23\ln\frac{m_b}{\bar{\mu}} 
    - 2\ln\frac{m_t}{\bar{\mu}}\Bigr)\Bigr\}^{-1}\biggr]\,.
\label{log}\end{align}

The last step is to match $\alpha_6$ directly to $\alpha_4$ 
(Fig.~\ref{fig:4} for $h=t,b$)
\begin{equation}
\alpha_6 \underset{\mathrm{LO}}{=}
   \alpha_4\!\bigm/\!\bigl[1 + (\alpha_4/3\pi)\bigl\{
      \ln(m_b/\bar{\mu}) + \ln(m_t/\bar{\mu})\bigr\}\bigr]\,,
\end{equation}
and eliminate $\alpha_6$ from (\ref{log}).  This yields
the complete set of LO logarithms
\begin{align}
\Gamma_{\mu 5}^{t,b}
\underset{\mathrm{LO}}{=}\,\frac{6\alpha_4}{23\pi}\gamma_\mu\gamma_5
\biggl[\Bigl\{1 + \frac{\alpha_4}{6\pi}
     \Bigl(25\ln\frac{m_b}{\bar{\mu}}\Bigr)\Bigr\}^{-1}&  \nonumber \\
   -\,\Bigl\{1 + \frac{\alpha_4}{6\pi}
     \Bigl( 23\ln\frac{m_t}{\bar{\mu}} 
         + 2\ln\frac{m_b}{\bar{\mu}}\Bigr)\Bigr\}^{-1}\biggr]&\,,
\end{align}
and hence the general asymptotic result \cite{BCST3}
\begin{align}  
\Gamma_{\mu 5}^{t,b}\, \sim\,
\frac{36}{23}\gamma_\mu\gamma_5\biggl[
 &\Bigl\{25\ln\frac{m_b}{\bar{\mu}}\Bigr\}^{-1} \nonumber \\
 -\ &\Bigl\{23\ln\frac{m_t}{\bar{\mu}}
        + 2\ln\frac{m_b}{\bar{\mu}}\Bigr\}^{-1}\biggr]\,.
\label{diag}\end{align}
This answer replaces (\ref{cancel}) and covers all cases:
\begin{enumerate}
\item It \emph{vanishes} for $m_t = m_b$.
\item It reproduces the correct sequential result (\ref{seq})
      when $t$ is decoupled and then $b$.
\item It works for intermediate cases as well:
      \[ 1 < \ln(m_t/\bar{\mu})/\ln(m_b/\bar{\mu}) < \infty \]
\end{enumerate}

\section{GENERALIZED RG}

Beyond LO, it is very difficult to identify all contributions 
``by hand'', so the RG becomes essential.  Our aim 
\cite{BCST1,BCST2,BCST3} is to extend the RG to take all large 
scale-invariant logarithms into account, and so generate asymptotic
expansions                           
for the simultaneous limit (\ref{1a}): LO, NLO (next to 
leading order), NNLO (next to NLO), and so on. Since no choice of scale 
can eliminate all large logarithms, we simply fix the 
$\overline{\mbox{\small MS}}$ scale $\bar{\mu}$ as $\{m_h\}\to\infty$, 
as Witten \cite{witten} did for a single heavy quark.

Witten has all asymptotic dependence on $m_h$ carried by 
a RG invariant running coupling $\widetilde{\alpha}_h$.  
{}For (say) $t$ decoupled from six-flavour QCD, 
$\widetilde{\alpha}_t$ is defined by the equation
\begin{equation}
\ln\bigl(m_t/\bar{\mu}\bigr)
=\, \int^{\run{t}}_{\alpha_6}\!\!dx\,\{1-\delta_6(x)\}/\beta_6(x)\,,
\label{top}\end{equation} 
where $\delta_F$ denotes the $F$-flavour Callan-Symanzik function 
for mass renormalization:
\begin{equation}
\delta_F(x) = - 2x/\pi + O(x^2)\,.
\label{g1}
\end{equation} 
Then $\alpha_6$ is eliminated by a matching condition \cite{BW} 
relating it to the residual coupling $\alpha_5$ which is held fixed as 
$m_t\to\infty$; (similarly, light-quark masses have to be matched).  This 
sequence can be continued by decoupling the $b$ quark
\begin{equation}
\ln\bigl({m_b}_5/\bar{\mu}\bigr)
= \int^{\run{b}}_{\alpha_5}\!\!dx\,\{1-\delta_5(x)\}/\beta_5(x) 
\end{equation} 
with $\alpha_5$ matched to $\alpha_4$, and so on. 

We begin by reformulating the matching procedure to make RG invariance
manifest. Once again, let $t$ be the heavy quark, and let $\bar{m}_t$
be its scale-invariant mass \cite{witten}:
\begin{equation}
\ln\bigl(\bar{m}_t/\bar{\mu}\bigr)
= \int_{\alpha_6}^{\run{h}}\hsp{-2.5}dx/\beta_6(x)\,.
\label{mass}\end{equation} 
Then RG invariance implies the existence of a \emph{matching function}
$\cal F$ defined by \cite{BCST2}
\begin{equation}
\ln\bigl(\bar{m}_t/\bar{\mu}\bigr)
= \int_{\alpha_5}^{\run{t}}\hsp{-2.5}dx/\beta_5(x) 
       + {\cal F}_{6\to 5}(\run{t})\,. 
\label{match}\end{equation} 
The result of eliminating $\run{t}$ from Eqs.~(\ref{mass}) and 
(\ref{match}) is the all-orders matching condition relating $\alpha_6$
to $\alpha_5$.  Similarly, there is a matching function $\cal G$ for 
light-quark masses.  For sequential decoupling, there corresponds a
sequence of matching functions
$({\cal F}_{6 \to 5},\ {\cal G}_{6 \to 5}),\ ({\cal F}_{5 \to 4},\ 
{\cal G}_{5 \to 4})$ and so on.  The $\cal F$ and $\cal G$ functions
are NNLO and NNNLO respectively.

Now consider a simultaneous limit, e.g.\ the decoupling of $t,b$. We
seek scale-invariant running couplings $\alpha_t$ and $\alpha_b$ 
which are defined \emph{without} reference to quantities $\alpha_5$
and ${m_b}_5$ associated with a $b$ threshold, but which tend to
Witten's couplings $\run{t}$ and $\run{b}$ in the sequential limit
(\ref{1c}). 

Our prescription \cite{BCST3} is to start with Witten's formula (\ref{top}) 
for the heaviest quark, and then use scale-invariant logarithms to generate
the others. For $t,b$ decoupling, we define $\alpha_t$ and $\alpha_b$
as follows,
\begin{gather} 
\ln\bigl(m_t/\bar{\mu}\bigr)
= \int^{\alpha_t}_{\alpha_6}\!\!dx\,
      \{1-\delta_6(x)\}/\beta_6(x)\ ,   \nonumber \\
\ln\bigl(m_t/m_b\bigr)
= \int^{\alpha_t}_{\alpha_b}\!\!dx\,
      \{1-\delta_5(x)\}/\beta_5(x)\,,
\label{def}\end{gather}
where both $m_t$ and $m_b$ are renormalized masses in the original 
\emph{six}-flavour theory. The integrands of Eq.~(\ref{def}) are
chosen such that, in the sequential limit, $\alpha_t$ and $\alpha_b$ 
agree with $\run{t}$ and $\run{b}$ to NLO accuracy.

The definitions of scale-invariant mass and matching functions also
have to be generalised. We define invariant masses $\overline{m}_t$
and $\overline{m}_b$ for the simultaneous limit by the formulas
\begin{gather}
\ln\bigl(\overline{m}_t/\bar{\mu}\bigr)
= \int^{\alpha_t}_{\alpha_6}\!\!dx/\beta_6(x)\ , \nonumber \\
\ln\bigl(\overline{m}_t/\overline{m}_b\bigr)
= \int^{\alpha_t}_{\alpha_b}\!\!dx/\beta_5(x)\,,
\label{x1}\end{gather}
and then construct matching functions $\cal F$ and $\cal G$ 
for $6\to 4$ flavours, e.g.
\begin{equation}
\ln\bigl(\overline{m}_b/\bar{\mu}\bigr)
= \int^{\alpha_b}_{\alpha_4}\!\!dx/\beta_4(x) 
  + {\cal F}_{6\to 4}(\alpha_t,\alpha_b)\,.
\label{x2}\end{equation}
The matching condition between $\alpha_6$ and $\alpha_4$ in the 
simultaneous $t,b$ limit is found by eliminating $\alpha_t$ and 
$\alpha_b$ from Eqs.~(\ref{x1}) and (\ref{x2}).

All $\{m_h\}$ dependence in the leading asymptotic power of an amplitude
is carried by the running couplings $\alpha_t,\,\alpha_b,\, \ldots$.
Thus, a power series expansion in  $\alpha_t,\,\alpha_b,\, \ldots$ 
yields coefficient amplitudes of the \emph{residual} theory:
\[ \mbox{Amplitude}\, =\, \sum_{k\ell\ldots}\alpha_t^k\alpha_b^\ell\ldots
                   {\cal A}_{k\ell\ldots}^{\mathrm{res}} \]
This expansion remains valid for the sequential limit with the 
\emph{same} coefficient amplitudes (to NLO --- beyond that, there are 
calculable corrections due to $\cal F$ and $\cal G$ functions).
Hence these coefficients can be found by matching against the result
for the sequential limit.

{}For $t,b$ decoupling from the weak neutral current, we found
the result \cite{BCST1,BCST3}
\begin{align}
\bar{t}\gamma_\mu\gamma_5&t - \bar{b}\gamma_\mu\gamma_5b\,
\underset{\mathrm{NLO}}{=}\, 
\frac{6}{23\pi}\bigl(\alpha_b-\alpha_t\bigr)   \nonumber \\
 \times&\Bigl\{1 + \frac{125663}{82800\pi}\alpha_b
          + \frac{6167}{3312\pi}\alpha_t\Bigr\}
             \Bigl(J_{\mu 5}^{\text{inv}}\Bigr)_4\,. 
\label{result}\end{align}
Notice that Eq.~(\ref{def}) implies $\alpha_t = \alpha_b$ for 
$m_t = m_b$, so both LO and NLO terms vanish in this case, as they 
should.  Our definitions imply
\begin{align}
\frac{6}{23\pi}\bigl(\alpha_b-\alpha_t\bigr)
\sim\ &\frac{36}{23}\biggl[
  \Bigl\{25\ln\frac{m_b}{\bar{\mu}}\Bigr\}^{-1}   \nonumber \\
 -\ &\Bigl\{23\ln\frac{m_t}{\bar{\mu}}
        + 2\ln\frac{m_b}{\bar{\mu}}\Bigr\}^{-1}\biggr]\,,
\end{align}
so the LO term in (\ref{result}) correctly reproduces the 
diagrammatic result (\ref{diag}).

The complete NLO result for the simultaneous decoupling of $t,b,c$
from $J_{\mu 5}^\mathrm{Weak}$ is given at the end of \cite{BCST1}.
In that case, the definition
\begin{equation}
\ln\bigl(m_b/m_c\bigr)
= \int^{\alpha_b}_{\alpha_c}\!\!dx\,
      \{1-\delta_4(x)\}/\beta_4(x) 
\end{equation}
is added to Eq.~(\ref{def}), and Eqs.~(\ref{x1}) and (\ref{x2}) are 
similarly amended so that $\alpha_6$ is matched directly to $\alpha_3$.


\end{document}